\documentstyle[aps,prl,epsfig,floats]{revtex}

\def\dert#1{{ d #1 \over d t}}

\newcommand{\nc}{\newcommand}
\nc{\postscript}[2] 
{\setlength{\epsfxsize}{#2\hsize}\centerline{\epsfbox{#1}}}
\nc{\bg}{B. Grzadkowski}
\nc{\non}{\nonumber}
\nc{\barx}{\bar{x}}\nc{\pbarn}{\;\hbox {pb}}\nc{\fbarn}{\;\hbox {fb}}
\nc{\vtrue}{v_0}
\nc{\vtree}{v}
\nc{\veff}{V_{\rm eff}}
\nc{\hc}{\hbox {h.c.}} 
\nc{\re}{\hbox {Re}} 
\nc{\im}{\hbox {Im}}
\nc{\mev}{\hbox {MeV}} 
\nc{\gev}{\;\hbox {GeV}} 
\nc{\tev}{\;\hbox {TeV}}
\def\gesim{\lower0.5ex\hbox{$\:\buildrel >\over\sim\:$}} 
\def\lesim{\lower0.5ex\hbox{$\:\buildrel <\over\sim\:$}} 
\nc{\xprd}[3]{{\it Phys.\ Rev.}\ {{\bf D{#1}} (#2), #3}}
\nc{\xprb}[3]{{\it Phys.\ Rev.}\ {{\bf B{#1}} (#2), #3}}
\nc{\xprl}[3]{{\it Phys.\ Rev.\ Lett.}\ {{\bf {#1}} (#2), #3}}
\nc{\pr}[3]{{\it Phys.\ Rep.}\ {{\bf {#1}} (#2), #3}}
\nc{\plb}[3]{{\it Phys.\ Lett.}\ {{\bf B{#1}} (#2), #3}}
\nc{\npb}[3]{{\it Nucl.\ Phys.}\ {{\bf B{#1}} (#2), #3}}
\nc{\ptp}[3]{{\it Prog.\ Theor.\ Phys.}\ {{\bf {#1}} (#2), #3}}
\nc{\zfp}[3]{{\it Z.\ Phys.}\ {{\bf C{#1}} (#2), #3}}
\nc{\mpla}[3]{{\it Mod.\ Phys.\ Lett.}\ {{\bf A{#1}} (#2), #3}}
\nc{\xrmp}[3]{{\it Rev.\ Mod.\ Phys.}\ {{\bf {#1}} (#2), #3}}
\nc{\ijmpa}[3]{{\it Int.\ J.\ Mod.\ Phys.}\
 {{\bf A{#1}} (#2), #3}}
\nc{\jhep}[3]{{\it JHEP}\ {{\bf #1} (#2), #3}}
\nc{\etal}{{\it et al.}}

\nc{\lspace}{\;\;\;\;\;\;\;\;\;\;} \nc{\llspace}{\lspace \lspace}

\nc{\beq}{\begin{equation}}  \nc{\eeq}{\end{equation}}
\nc{\bea}{\begin{eqnarray}}  \nc{\eea}{\end{eqnarray}}
\nc{\baa}{\begin{array}}     \nc{\eaa}{\end{array}}
\nc{\bit}{\begin{itemize}}   \nc{\eit}{\end{itemize}}
\nc{\ben}{\begin{enumerate}} \nc{\een}{\end{enumerate}}
\nc{\bce}{\begin{center}}    \nc{\ece}{\end{center}}

\nc{\mh}{m_h}
\nc{\mt}{m_t}
\nc{\mz}{m_Z}
\nc{\la}{\lambda}
\nc{\La}{\Lambda}

\def\half{\frac12}
\def\lcal{{\cal L}}
\def\up#1{^{(#1)}}
\def\inv#1{\frac1{#1}}
\def\ocal{{\cal O}}
\def\pb{\bar\varphi}

\begin{document}

\def\mib#1{\mbox{\boldmath $#1$}}
\def\bra#1{\langle #1 |} \def\ket#1{|#1\rangle}
\def\vev#1{\langle #1\rangle} \def\dps{\displaystyle}

\twocolumn[\hsize\textwidth\columnwidth\hsize\csname @twocolumnfalse\endcsname


\begin{flushright}
IFT-18-01\\
UCRHEP-T311\\
June, 2001
\end{flushright}

\title{\Large\bf Bounds on the Higgs-Boson Mass\\
 in the Presence of Non-Standard Interactions}

\renewcommand{\thefootnote}{\alph{footnote})}
\author{\sc Bohdan Grzadkowski} 
\address{Institute of Theoretical Physics, Warsaw University,
 Ho\.za 69, PL-00-681 Warsaw, POLAND\\
 E-mail: {\tt bohdan.grzadkowski@fuw.edu.pl}}

\author{\sc Jos\'e Wudka}
\address{Department of Physics, University of California, Riverside CA
 C 92521-0413, USA\\
E-mail: {\tt jose.wudka@ucr.edu}}


\maketitle

\begin{abstract}

The triviality and vacuum stability bounds on the Higgs-boson mass are revisited
in the presence of new interactions parameterized
in a model-independent way by an effective lagrangian.
When the scale of new physics $\La$ is below $ 50 \tev$
the triviality  bound is unchanged but
the  stability lower bound is increased
by $40\div60\gev$. Should the Higgs-boson mass be close to its current
lower experimental limit, this leads to the possibility of new physics at the
scale of a few$\tev$, even for modest values of the effective lagrangian
parameters.
\end{abstract} 

\draft

\vskip1pc]

\paragraph{Introduction}
\label{sect:intro}

In spite of a huge experimental effort, the Higgs particle, the last missing ingredient of 
the Standard Model (SM) of electroweak interactions has not been discovered yet. 
For a Higgs-boson mass $\mh \lesim 115 \gev$ the most promising production channel
at LEP2 has been $e^+e^- \to Z h$ that provides the limit~\cite{higgs_limit} 
$ \mh > 113.2 \gev$ for the SM Higgs-boson.
The Higgs particle also contributes radiatively to several well measured
quantities and this can be used to derive the
upper bound~\cite{prec_data} $\mh \lesim 212 \gev$ at 95 \% C.L.. 
Both these limits are, however, highly model-dependent.

There also exist  theoretical constraints on $\mh$ based on the so-called
triviality and vacuum stability arguments. As it is well know~\cite{triviality} 
the renormalized $\phi^4$ theory cannot contain an interaction term ($\la \phi^4$) for
any non-zero scalar mass: the theory must be trivial. Within a 
perturbative approach this statement corresponds to the fact
that the running coupling constant $\la(\kappa)$ necessarily diverges 
at a finite value of the renormalization scale $ \kappa $
(the Landau pole). Consequently, only a non-interacting
theory is consistent at all energy scales. 
An analogous effect occurs in the scalar sector of the SM 
(modified to some extend by the presence
of gauge and Yukawa interactions). This, however,
does not necessary imply a trivial
scalar sector since there is no reason
to believe the SM to be valid at arbitrarily high energy scales.

Assuming the SM is an effective theory applicable only below an energy
scale $ \La$, the Landau pole
should occur at scale $ \La $ or above, and this condition gives a
($\La$-dependent) upper bound on $ \mh $~\cite{triv_bounds}. 
On the other hand, for sufficiently small $ \mh $ radiative corrections can
destabilize the ground state. Then  the requirement that the 
SM vacuum is stable for scales below $ \La$ provides a
lower bound on $ \mh $~\cite{vacuum_bounds}. 

The triviality and vacuum stability bounds on $\mh$ are 
usually derived assuming SM interactions. However,
if the scale of new physics is sufficiently low ($\sim$ a few TeV)
one could expect non-standard interactions to generate important modifications
of these constraints.
The problem deserves special attention given the  possibility
that the Higgs-boson was observed at LEP2~\cite{higgs_disc}
with a mass $\mh\simeq 115 \gev$; in this case
the SM constraint from vacuum stability
requires $ \Lambda \lesim {\cal O}(100) \tev $~\cite{quiros} 
(the precise number depends on the
top quark mass), allowing for the attractive possibility
that $ \La$ is actually much lower, even at the 
level of a few TeV.

In this letter we determine the manner in which heavy
physics with scales in the $ 10 \tev $ region modify the theoretical
constraints on the Higgs-boson mass. We follow a model-independent
approach parameterizing the heavy physics
effects by an effective Lagrangian satisfying the SM gauge
symmetries. Then, using standard procedures, we derive the stability and
triviality bounds on $ \mh $ as a function of the heavy physics scale $
\La $.
Since LHC, the future proton-proton collider, is expected to be sensitive to 
scales $ \La $ of the order of a few TeV, the results will be presented
for scales between $ 0.5 $ and $ 50 \tev$.

\paragraph{Non-Standard Interactions}
\label{non_stand}

Our study of the stability and triviality constraints on the Higgs-boson mass
will be based on the SM Lagrangian modified by the addition of a series
of gauge-invariant
effective operators $ \ocal_i $ whose coefficients 
$\alpha_i $ parameterize the low-energy 
effects of the heavy physics~\cite{leff.refs}. Assuming that these non-standard effects
decouple implies~\cite{decoupling} that these operators appear multiplied
by appropriate inverse powers of $ \La $. The leading effects are then
generated by operators of mass-dimension 6\footnote{Dimension 5 operators
violate lepton number~\cite{effe_oper} and are associated with new physics at
very large scales, so we can safely ignore their effects.}.
Given our emphasis on Higgs-boson physics the effects of all fermions
excepting the top-quark can be ignored~\footnote{We assume that 
chirality-violating effective interactions are natural~\cite{thooft},
being suppressed by the corresponding Yukawa couplings.}. We then have
\bea
&& \lcal_{\rm tree} = 
-\frac14 ( F^2 + B^2) 
+\left| D \phi \right|^2 
+i \bar q \not\!\!D q
+i \bar t \not\!\!D t 
\cr
&& \qquad + f \left( \bar q \tilde\phi t + \hbox{h.c.} \right)
- \lambda \left(|\phi|^2 -\vtree^2/2\right)^2
+ \sum_i\frac{\alpha_i}{\La^2}{\cal O}_i,
\label{lagrangian}
\eea
where $\phi$ ($\tilde \phi = -i \tau_2 \phi^* $), 
$q$ and $t$ are the scalar doublet, third generation 
left-handed quark doublet and the right-handed top singlet, respectively. 
$D$ denotes the covariant derivative, $F_{\mu\nu}^i$ and $B_{\mu\nu}$ 
the $SU(2)$, $U(1)$ field strengths
whose couplings we denote by $g$ and $g'$.

When the heavy interactions are weakly coupled the leading effects
at low energy are determined by those $ \alpha_i $ generated
at tree level by the heavy physics 
(loop-generated coefficients are suppressed by 
coupling constants and numerical 
factors $ \sim 1/ (4\pi)^2 $~\cite{tree_oper}). Because of this we will consider only those
operators that can be generated at tree-level by the heavy physics. 
There are 81 dimension-six operators (for one family)~\cite{effe_oper},
but only 16 can be generated at tree-level and involve only the fields in
(\ref{lagrangian}). Of these
5 contribute directly to the effective potential, the remaining 11
affect the results only through their RG mixing and, being suppressed by
a factor $ \sim 1/(G_F\La^2) $, will play a sub-dominant role.
We will include only one of these operators
to illustrate these effects.

In the calculations below we will include the set
\beq
\baa{lll}
\!\!\!{\ocal_{\phi}} = \inv3 | \phi|^6 &
\!\!\!{\ocal_{\partial\phi}} = \half \left( \partial | \phi |^2 \right)^2 &
\!\!\!\!\!\!\!\!\!\!\!{\ocal_{\phi}\up1} = | \phi |^2 \left| D \phi \right|^2 \cr
\!\!\!{\ocal_{\phi}\up3} = \left| \phi^\dagger D \phi \right|^2 &
\!\!\!{\ocal_{t\phi}} = | \phi |^2 (\bar q \tilde\phi t + \hbox{\small{h.c.}}) &
\!\!{\ocal_{qt}\up1} = \half \left|\bar q t \right|^2
\label{operators}
\eaa
\eeq
where the first 5 operators contribute directly to the effective
potential, while ${\ocal_{qt}\up1}$ is included to estimate the effects
of RG mixing. Note that only $\ocal_{\phi} $ contributes
at the tree level to the scalar potential:
\beq
V\up{\rm tree}= 
- \eta \Lambda^2 |\phi|^2 + \lambda |\phi|^4 - 
{\alpha_{\phi} \over 3 \Lambda^2 } | \phi|^6
\label{tree_pot}
\eeq
where $\eta \equiv \lambda v^2/\Lambda^2 $.

\paragraph{Triviality Bound}
\label{triv_bound}

In order to study the high energy behavior of the scalar potential we derive
the RG running equations for $\la$, $\eta$ and the $\alpha_i$.
This running is
also influenced by the gauge and Yukawa interactions, so the full RG evolution
requires the $ \beta $ function for all these couplings.
Using dimensional regularization in the $\overline{\rm MS}$ 
scheme, and defining $ \bar\alpha = \alpha_{\partial\phi} +2 
\alpha_\phi\up1 + \alpha_\phi \up3 $ we find: 
\begin{eqnarray}
\dert \lambda &=&12\lambda^2 -3 f^4 + 6 \lambda f^2 -(3\la/2)\left(3 g^2 + g'{}^2 \right) 
\cr && 
+(3/16) \left(g'{}^4 +2 g^2 g'{}^2 + 3 g^4\right) 
\cr &&
- 2 \eta \left[2 \alpha_\phi + 
\lambda \left( 3 \alpha_{\partial\phi} +4 \bar\alpha + \alpha_\phi\up3\right) \right] 
\cr
\dert \eta &=& 3\eta\left[2\lambda + f^2 - \left( 3g^2+ g'{}^2 \right)/4\right]
-2 \eta^2  \bar\alpha
\cr
\dert f &=& 9 f^3/4 -(f/2)\left(8 g_s^2 +  9 g^2/4 + 17 g'{}^2/12 \right) \cr
&& + 3 \eta\alpha_{t\phi} 
- (f\eta/2) \left(\bar \alpha + 3 \alpha_{qt}\up1 \right)  \cr
\dert{\alpha_\phi}&=& 
 9 \alpha_\phi \left(6 \lambda + f^2 \right)
 + 12 \lambda^2 (9\alpha_{\partial\phi}+6 \alpha_\phi\up 1
 +5\alpha_\phi\up 3)
\cr &&  \hspace{-10pt} + 36 \alpha_{t\phi} f^3 - (9/4) ( 3 g^2 + g'{}^2) \alpha_\phi 
\cr && 
 \hspace{-10pt} - (9/8)\left[2  \alpha_\phi\up1 g^4 + \left(\alpha_\phi\up1 + 
\alpha_\phi\up3 \right)(g^2 + g'{}^2 )^2 \right] \cr
\dert{\alpha_{\partial\phi}}&=& 2 \lambda
 \left( 6 \alpha_{\partial\phi} - 3\alpha_\phi\up1 +\bar\alpha \right)
 + 6 f \left(f \alpha_{\partial\phi}  - \alpha_{t\phi}\right) \cr
\dert{\alpha_\phi\up1}&=& 2 \lambda 
 \left(\bar\alpha+3\alpha_\phi\up1\right)
 + 6 f \left(f \alpha_\phi\up1 
 -\alpha_{t\phi}\right) \cr
\dert{\alpha_\phi\up3}&=& 6 (\lambda +f^2) \alpha_\phi\up3 \cr
\dert{\alpha_{t\phi}} &=& -3 f (f^2+\lambda) \alpha_{qt}\up1 
+ (15f^2/4-12\lambda) \alpha_{t\phi} \cr &&
- (f^3/2) \left( 
\alpha_{\partial\phi} - \alpha_\phi\up1 + \bar \alpha \right) \cr
\dert{\alpha_{qt}\up1} &=&(3/2) \alpha_{qt}\up1 f^2
\label{beta_fun}
\end{eqnarray}
where we neglected terms quadratic in the $ \alpha_i $, and
$t\equiv \log (\kappa/\mz)/(8\pi^2)$,
$ \kappa$ being the renormalization scale. The evolution 
of the gauge couplings
$g$, $g'$ and $ g_s$ (for the strong interactions) are unaffected
by the $\alpha_i$.

In order to solve the equations (\ref{beta_fun}) we have to specify
appropriate boundary conditions.
For the SM parameters these are determined by requiring that the correct
physical parameters 
are obtained at the electroweak scale. The values of $ \lambda,~ f$ and $\eta$
at $t=0$ are fixed using the physical Higgs-boson mass~\cite{quiros}, the top mass
and the scalar field vacuum expectation value.
Since the experimental errors in the top-quark mass are larger  than the
expected deviations from the tree-level expression we  
use $\mt= \vtrue f/ \sqrt{2} = f\times 174\gev$ for simplicity. 
We also require that the solutions to (\ref{beta_fun}) reproduce
the correct electroweak vacuum where the scalar field has the
expectation value 
$ \langle \pb \rangle \simeq \vtrue/\sqrt{2}$~\footnote{
We ignored a small correction to the $W$ mass 
$\propto  \alpha_\phi\up1 (\vtrue/\La)^2 $ which modifies the relationship
between $G_F$ and $ \vtrue$ at the 10\%~level; changing
the bounds on $\mh $ by $ \lesim 6\%$.}.
The relation between the SM tree-level vacuum
$\vtree$ and the physical electroweak vacuum $\vtrue $ is
\beq
\vtrue =\vtree
- {1 \over 4 \la(0) \vtrue ^2}
\left.{\partial (\veff-V_{\rm SM}^{\rm (tree)}) 
\over \partial( \pb/ \sqrt{2} )}\right|_{\pb={\vtrue / \sqrt{2}}},
\label{vev}
\eeq
where $V_{\rm SM}^{\rm (tree)}$ denotes the tree-level SM potential, $\veff$
the effective potential calculated up to 1-loop including all
effective operator contributions, and $ \la(0) $ the
running coupling constant evaluated at $ t=0 $.
Finally we require that the gauge coupling constants satisfy 
$ g(0) = 0.648,~ g'(0)= 0.356,~ g_s(0)=1.218$.

\begin{figure}[htb]
\psfull
\begin{center}
  \leavevmode
\epsfig{file=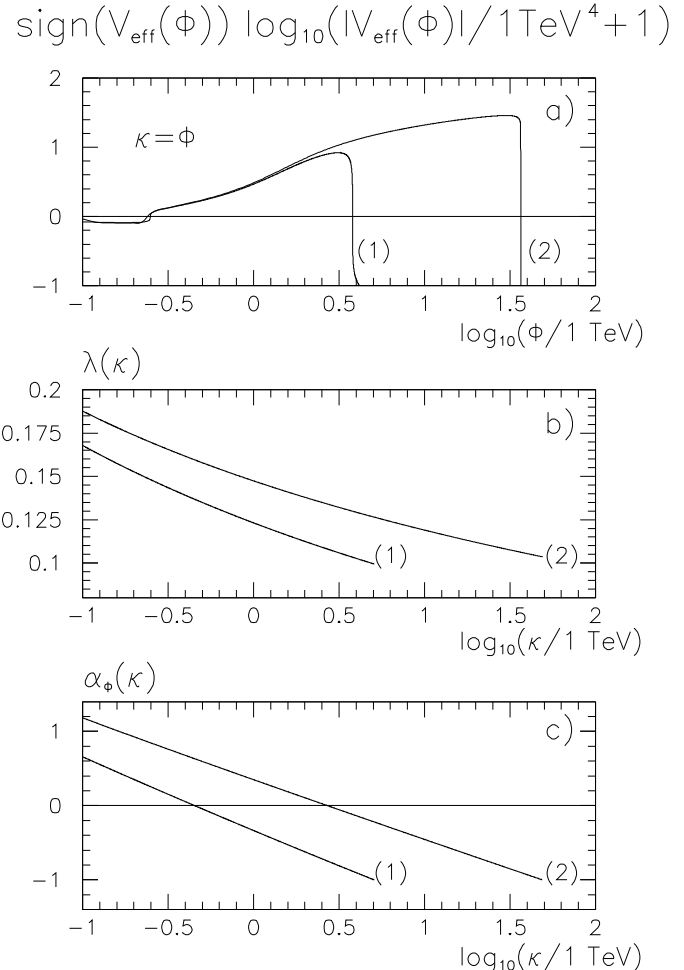,height=13cm,width=15cm,bbllx=4cm,bblly=1.2cm,bburx=26cm,
bbury=20cm,angle=0}
 \end{center}
\caption{\emph{The effective potential renormalized at the scale $\kappa=\phi$ (a);
(we have plotted 
${\rm sgn}(\veff)\log_{10}[(\veff/1\tev^4)+1]$
in order to make visible the shallow minimum at $\pb=
\vtrue/\sqrt{2}$). 
The running of $\lambda$ (b) and $\alpha_\phi$ (c) when $ \alpha_i(\La)=-1$,
$\mt = 175 \gev $, for 
$ \La =5.1 \tev$, $\mh=140.4 \gev $ (curves (1)) and $ \La = 48.9
\tev$, $\mh = 148.7 \gev $ (curves (2)).
}}
\label{potential}
\end{figure}

The boundary conditions for the $\alpha_i$ are naturally specified at the scale $\kappa=\La$. 
For the weakly-coupled heavy interactions considered here it is natural
to assume that $\alpha_i|_{\kappa=\La} \lesim {\cal O}(1)$~\cite{tree_oper}
(the triviality bounds are insensitive to the precise value). In Fig. \ref{potential}
(b),(c) we have plotted two examples of the solutions to (\ref{beta_fun}) using the above
boundary conditions.

\begin{figure}[htb]
\psfull
\begin{center}
  \leavevmode
\epsfig{file=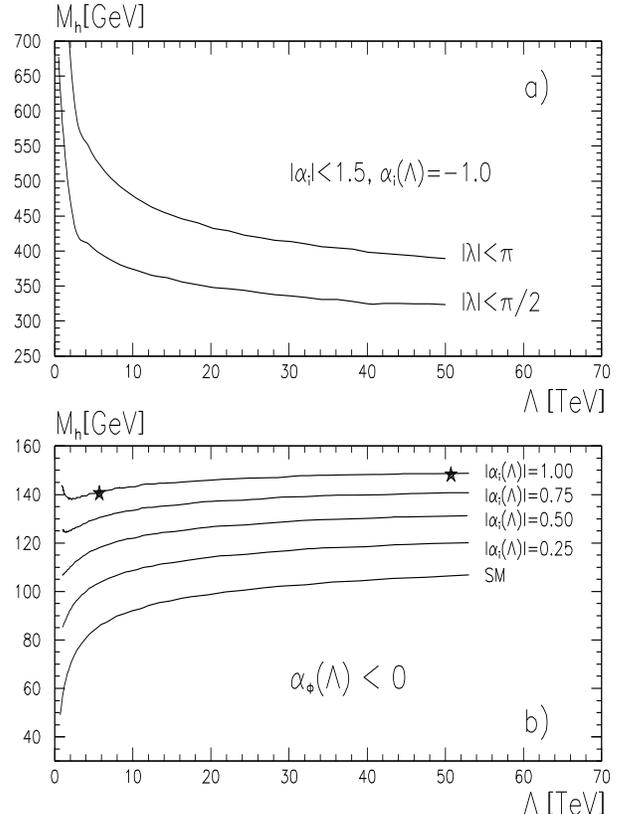,height=10cm,width=10cm,bbllx=0cm,bblly=1.3cm,bburx=22cm,
bbury=18cm,angle=0}
 \end{center}
\caption{\emph{(a) Triviality bound on the Higgs-boson mass obtained from
(\ref{triv_con}). (b) The stability bound on $ \mh$ obtained from
(\ref{stab_con}) when $ \alpha_\phi(\Lambda) < 0 ,~\mt=175\gev $.
The  stars correspond to the solutions (1) and (2)
of Fig. \ref{potential}.
}}
\label{bounds}
\end{figure}

The triviality
bound on $\mh$ will be obtained by requiring $\la$ and $\alpha_\phi$ to remain
below specified values (as opposed from requiring an actual divergence)
\beq
\la(t)\le \la_{\rm max}\quad
|\alpha_i(t)|\le \alpha_{\rm max}
\label{triv_con}
\eeq
for all scales below $\La$.
We will present results for $\alpha_{\rm max}=1.5$ and
$ \lambda_{\rm max} = \pi $ and $ \pi/2 $ (this result
is insensitive to the choice of $ \alpha_{\rm max} $). The bound
obtained by saturating either of the above inequalities 
is plotted in  Fig. \ref{bounds} (a). We note
that the results are almost identical to the ones obtained for the
SM~\cite{triv_bounds}.

In order to understand qualitatively the absence of significant
corrections to the triviality bound, it is useful to switch off all 
$\alpha_i$ except $\alpha_\phi$. Then, since
$ d \ln(\alpha_\phi)/dt>0 $\footnote{
Here we consider heavy Higgs bosons, therefore $\la$ 
remains positive in the whole integration region.},
$| \alpha_\phi| $ will decrease when evolving 
from the scale $\La$ downwards, reaching values 
$\sim 10^{-1}\div 10^{-2}$ at $\kappa=\mz$  (for $ |\alpha_\phi(\La) |\sim 1 $). 
In addition, note also that the effects of $ \alpha_\phi $ on
the evolution of the SM parameters are suppressed by a
factor of $ \eta $. Hence the effects of $ \alpha_\phi $ on
the RG evolution of $ \la$ are screened, leaving the SM triviality bound
essentially unaffected.

\paragraph{Vacuum Stability Bound}
\label{vac_bound}

In order to investigate the vacuum structure of the effective theory we will
first calculate the effective potential:
\begin{equation}
\veff(\pb) = - \sum_n \inv{n!} \Gamma\up n(0) \pb^n,
\label{eff_pot_def}
\end{equation}
where $\Gamma\up n(0)$ are n-point 1PI vertices at
zero external momenta and $\pb$ is the classical scalar field.
Using the Landau gauge\footnote{The loop contributions to $ \veff $ are
gauge dependent~\cite{gauge_dep_eff_pot}, yet since the RG-improved 
tree-level effective potential is gauge-invariant, the stability bound
depends weakly on the gauge parameter leading to an uncertainty
$\Delta \mh \lesim .5 \gev$~\cite{quiros}.} 
we find:
\begin{eqnarray}
\veff(\pb) &=& V\up{\rm tree}
+ {1 \over 64 \pi^2} \sum_{i=0}^5 c_i R_i^2 [\ln (R_i/\kappa^2)-\nu_i]
\label{effpot}
\end{eqnarray}
where $c_0=-4,~c_1=1,~c_{2,4}=3,~c_3=6,~c_5=-12$, $\nu_{0,1,2,5}=3/2,~\nu_{3,4}=5/6$, 
$ R_0 = \eta\La^2 $ and
\begin{eqnarray}
R_1 &=& \lambda (6 |\pb|^2 -\vtree^2 )\left[1 -
(2 \alpha_{\partial\phi} + \alpha_\phi\up1 + \alpha_\phi\up3 )|\pb|^2/\La^2\right]
 \cr && 
- 5 \alpha_\phi |\pb|^4/\Lambda^2 \cr
R_2 &=& \lambda (2 |\pb|^2  -\vtree^2 ) \left[1-(
\alpha_\phi\up1 + \alpha_\phi\up3/3)|\pb|^2/\La^2\right]
\cr && 
 - \alpha_\phi  |\pb|^4/ \Lambda^2 \cr
R_3 &=& (g^2/2)|\pb|^2 \left( 1 +  |\pb|^2 \alpha_\phi\up1/ \Lambda^2  \right) \cr
R_4 &=& [( g^2 +g'{}^2)/2] |\pb|^2 \left( 1 +  |\pb|^2 ( \alpha_\phi\up1 + 
\alpha_\phi\up3) / \Lambda^2  \right) \cr
R_5 &=& f |\pb|^2 \left(f + 2 \alpha_{t \phi} | \pb|^2 / \Lambda^2 \right),\non
\end{eqnarray}
 It is understood that the above expression is
accurate up to corrections of order $ 1 / \La^4 $.

The {\em form} of the effective potential is precisely the same as the
one in the pure SM, the whole effect of the effective operators can be
absorbed in a re-definition of the $R_i$. We also include the
scale dependence of $ \pb $: 
$ 
\pb(t)=\exp\left\{-\int_0^t 
\gamma dt'\right\} \pb(t=0),
$ where
$\gamma = 3 f^2/2 - 3(3g^2+ g^{\prime 2})/8
- \eta\bar\alpha/2$
(the couplings appearing in $ \gamma $ are understood to be the
solutions to (\ref{beta_fun})). In the following we
will consider the RG improved effective potential $\veff( \pb(t)) $
defined using (\ref{effpot}) and $ \pb(t)$.

We note that $\veff( \pb(t)) $
is scale invariant, that is, $ \kappa \; d \veff(\pb(t))/d\kappa =0 $ 
(to one loop and ignoring terms quadratic in the $ \alpha_i$).
In verifying this relation
the constant term in (\ref{effpot}) must be chosen appropriately, 
our choice is determined
by the requirement  $\veff(\pb=0)=0$, which is consistent with (\ref{eff_pot_def});
for details see Ref.~\cite{cosm_const}.
 
When using the above expressions to derive the stability bound on $ \mh
$ we will need to consider values of $\pb$ substantially larger then the electroweak
scale $\vtrue $. Therefore we shall choose a renormalization scale 
$\kappa \sim \pb$ in order
to moderate the logarithms that appear in $\veff$.

Fig. \ref{potential} illustrates the behaviour of the effective potential
renormalized at the scale $\kappa=\bar{\varphi}$. 
To show the relevance of RG running of effective-potential parameters
we also plot the evolution of $\la$ and $\alpha_\phi$
for two sets of initial conditions
corresponding to the $(\mh,\La)$ values
marked in Fig. \ref{bounds} by stars. As it is seen from the figure effects
of the running are substantial, e.g. for the 
set (2) $\la$ changes by almost 100\%
while $\alpha_\phi$ by more than 200\%\ and undergoes a sign change. This
emphasises the fact that the RG running of the coefficients $\alpha_i$
must be included when studying 
the vacuum stability of the system\footnote{Some corrections to the 
SM vacuum stability bound due to
 $\ocal_\phi$ were discussed in~\cite{datta}.
However, the authors did not consider one-loop contributions to
the effective potential generated by $ \ocal_\phi$, nor the
RG running of $\alpha_\phi$.}.

The initial conditions for the running couplings
guarantee that the electroweak vacuum is 
at $\langle \pb \rangle=\vtrue /\sqrt{2}$. 
However if $\veff $ at some large value of the field $\pb_{\rm high} $
is smaller than $ \veff(\langle \pb \rangle)$
this vacuum becomes unstable (as there would be a possibility of 
tunnelling\footnote{The question of the tunnelling 
rate will not be
discussed in this latter, we shall assume 
that tunnelling is possible with the transition
time smaller than the age of the universe.} towards
the region of lower energy).
This will occur when the Higgs-boson mass is sufficiently small
(corresponding to a small value of $ \lambda(0)$), and provides a lower
bound on $ \mh$. In this case $ \pb_{\rm high} $
defines a scale at which the theory breaks down, so that $ \pb_{\rm high} \sim
\Lambda $. In actual calculations we took $ \pb_{\rm high} = 0.75 \Lambda $ 
since (\ref{lagrangian}) is valid for scales below $ \Lambda $, hence
the stability bound on $ \mh$ is determined by the condition
\beq
\veff(\pb =0.75\La)|_{\kappa=0.75\La} =
\veff(\pb = \vtrue /\sqrt{2})|_{\kappa=\vtrue /\sqrt{2}}
\label{stab_con}
\eeq
where, as mentioned previously, we have chosen the renormalization scale
$\kappa$ to tame the effects of the logarithmic contributions to
$\veff(\pb)$. The resulting bound on $ \mh $ as a function of $ \La$
for various choices of $ \alpha_i(\La) $ is plotted in
Fig. \ref{bounds} (b). 

In obtaining the stability bounds of Fig. \ref{bounds} (b) we assumed all
couplings $ \alpha_i $ had the same magnitude at the high scale $
\Lambda $, and $ \alpha_\phi< 0$ (the results are insensitive to the 
sign of the other $ \alpha_i $). For other values of $\alpha_i$ we
found that when $\La > 300\gev$ there is a curve in the $\alpha_{\phi} - \alpha_{t\phi}$
plane below which  either $\pb=174\gev$ is not a minimum or, if it is,
then there is another deeper minimum at a scale
$174 \gev < \pb<\La$; we can roughly say
that this unphysical scenario can be avoided if
$\alpha_\phi \lesim -0.1$ ~\footnote{We do not
expect this result to be modified significantly when terms of order
 $ 1/\La^4 $ are included: a contribution $ \sim \alpha\up 8
\pb^8/\Lambda^4 $ can balance the destabilizing effect of $ \ocal_\phi $
only when $ \pb \sim \La $ which again leads to $ \La \sim 300 \gev $.}.

There is an important remark here in order. 
The SM vacuum stability bound together with the
experimental limit $ \mh > 113.2 \gev$ implies $ \Lambda \lesim {\cal O}(100) \tev $.
Assuming now that the limit on $ \mh $
remains unchanged in presence of effective operators~\cite{j_b,datta},
then, as seen from Fig. \ref{bounds} (b), 
even for the modest values $|\alpha_i|=0.25,~0.50,~0.60$ the upper bound on
$ \La$ is significantly reduced to   $\La \simeq 20,~4~1\tev$, 
respectively!

\def\maybe{Other limits on the scale $\La$ could be obtained form so called precision observables. 
The most elegant approach is
to calculate the oblique parameters $S$, $T$ and $U$~\cite{stu} within the effective 
theory\footnote{It should be noticed that among operators considered 
here only $\ocal_\phi^{(3)}$
contributes directly to the oblique parameters ($T$) 
and therefore is constrained by the precision data, however as it has been 
shown here the operator that is most relevant for the triviality and vacuum stability bound is
$\ocal_\phi$ and contributions from $\ocal_\phi^{(3)}$ are much less important.}
and then fit their experimental values~\cite{stu_effective_1,stu_effective_2}. 
The limits obtained that way depend also on the Higgs-boson
mass $\mh$ therefore it would be interesting to superimpose precision-measurement limits,
the direct LEP limit and 
those obtained here, consistently taking into account higher dimensional operators,
that is however beyond the scope of this paper\footnote{Searches that neglect 
higher-dimensional-operator
corrections to both the triviality and the vacuum stability Higgs-boson bounds 
are published, see Ref.\cite{stu_effective_2}.}. }

\paragraph{Summary and Conclusions}

We have considered the triviality and stability
restrictions on $\mh$ when the SM is the low-energy limit
of a weakly-coupled decoupling theory of typical scale $\La$.
It was assumed that there is a significant gap between $\La$ and the
typical experimental energies so that the heavy interactions can 
be accurately described by a set of effective vertices. 
We showed that for the scale of new physics in the region 
$0.5 \tev \le \La \le 50 \tev$
the SM triviality (upper) bound remains unmodified. In contrast
stability (lower) bound  could be increased
by $40-60\gev$ reducing substantially the allowed region of $\mh $ 
values. If $\mh$
is close to its current experimental limit, then the maximum allowed value of $\La$
would be decreased dramatically even for modest values of coefficients of effective 
operators, implying new physics already at the scale of a few TeV.

\acknowledgements
This work is supported in part by the State
Committee for Scientific Research (Poland) under grant 5~P03B~121~20
and funds provided by the U.S. Department of Energy under grant No.
DE-FG03-94ER40837. One of the authors (BG) is indebted to
CERN, SLAC and U.C. Riverside for the
warm hospitality extended to him while this 
work was being performed.


\end{document}